\newtheorem{theorem}{Theorem}
\newtheorem{definition}{Definition}

\documentclass[12pt]{article}%
\usepackage{amssymb}
\usepackage{amsmath}
\usepackage{amsfonts}
\usepackage{graphicx}%
\begin{document}

\title{COMPLEXITY IN QUANTUM SYSTEM AND\ ITS\ APPLICATION TO\ BRAIN\ FUNCTION}
\author{Masanori Ohya\\Department of Information Sciences,\\Science University of Tokyo\\278 Noda City, Chiba, Japan}
\date{}
\maketitle

\section{Introduction-What is complexity?-}

The complex system has been considered by Casity in Santafe research center as follows:

(1) A system is composed of several elements called agents. The size of the
system (the number of the elements) is medium.

(2) The agent has intellegence.

(3) Each agent has interaction due to local information. The decision of each
agent is determined by not all information but the limited information of the system.

Under a small modification, I define the complex system as follows:

(1) A system is composed of several elements. The scale of the system is often
large but not always, in some cases one.

(2) Some elements of the system have special (self) interactions (relations),
which produce a dynamics of the system.

(3) The system shows a particular character (not sum of the characters of all
elements) due to (2).

\begin{definition}
A system having the above three properties is called ``complex system''. The
''complexity''of such a complex system is a quantity measuring that
complexity, and its change describes the appearance of the particular
character of the system.
\end{definition}

There exist such measures describing the complexity for a system, for
instance, variance, correlation, level - statistics, fluctuation, randomness,
multiplicity, entropy, fuzzy, fractal dimension, ergodicity (mixing, flow),
bifurcation, localization, computational complexity (Kolmogorov's or
Chaitin's), catastrophy, dynamical entropy, Lyapunov exponent, etc. These
quantities are used case by case and they are often difficult to compute.
Moreover, the relations among these are lacking (not clear enough). Therefore
it is important to find common property or expression of these quantities. In
this paper, we introduce such a common degree to describe the chaotic aspect
of quantum dynamical systems. Further we describe the function of barin in the
framework of information dynamics \cite{O3,O12}(ID for short) and we discuss
the value of information attached to the brain in terms of the complexity in
ID and the chaos degree\cite{O7,O9,O10}.

\section{Quantum Information Dynamics}

There are two aspects for the complexity, that is, the complexity of a state
describing the system itself and that of a dynamics causing the change of the
system (state). The former complexity is simply called the ''complexity'' of
the state, and the later is called the ''chaos degree'' of the dynamics in
this paper. Therefore the examples of the complexity are entropy, fractal
dominion, and those of the chaos degree are Lyapunov exponent, dynamical
entropy, computational complexity. Let us discuss a common quantity measuring
the complexity of a system so that we can easily handle. The complexity of a
general quantum state was introduced in the frame of ID \cite{O3,IKO,O12} and
the quantum chaos degree was defined in \cite{IKO2}, which we will review in
this section.

Information Dynamics is a synthesis of dynamics of state change and complexity
of state. More precisely, let $(\mathcal{A},\mathfrak{S},\alpha(G))$ be an
input (or initial) system and $(\overline{\mathcal{A}},\overline
{{\mathfrak{S}}},\overline{\alpha}(\overline{G}))$ be an output (or final)
system. Here $\mathcal{A}$ is the set of all objects to be observed and
$\mathfrak{S}$ is the set of all means for measurement of $\mathcal{A}$,
$\alpha(G)$ is a certain evolution of system. Once an input and an output
systems are set, the situation of the input system is described by a state, an
element of $\mathfrak{S}$ $,$ and the change of the state is expressed by a
mapping from $\mathfrak{S}$ to $\overline{{\mathfrak{S}}}$, called a channel,
$\Lambda^{*}:\mathfrak{S}\to\overline{{\mathfrak{S}}}$ . Often we have
$\mathcal{A}=\overline{\mathcal{A}}$, $\mathfrak{S}=\overline{{\mathfrak{S}}}%
$, $\alpha=\overline{\alpha},$ which is assumed in the sequel. Thus we claim

\begin{center}
[Giving a mathematical structure to input and output triples

$\equiv$ Having a theory]
\end{center}

\smallskip For instance, when $\mathcal{A}$ is the set $M(\Omega)$ of all
measurable functions on a measurable space $(\Omega,\mathcal{F})$ and
$\mathfrak{S}(\mathcal{A})$ is the set $P(\Omega)$ of all probability measures
on $\Omega$ , we have usual probability theory, by which the classical
dynamical system is described. When $\mathcal{A}$ = $B(\mathcal{H}),$ the set
of all bounded linear operators on a Hilbert space $\mathcal{H}$, and
$\mathfrak{S}(\mathcal{A})$ = $\mathfrak{S}(\mathcal{H})$ , the set of density
operators on $\mathcal{H}$, we have a usual quantum dynamical system. In this
paper, we assume that both the input and output triple $(\mathcal{A}%
,\mathfrak{S},\alpha(G))$ is a C*-dynamical system or the usual quantum system
as above, and a channel, $\Lambda^{*}:\mathfrak{S}\to\mathfrak{S}$ is a
completely positive map.

There exist two complexities in ID, which are axiomatically given as follows:

Let $(\mathcal{A}_{t},\mathfrak{S}_{t},\alpha^{t}(G^{t}))$ be the total system
of both input and output systems; $\mathcal{A}_{t}\equiv\mathcal{A\otimes A}
,\mathfrak{S}_{t}\equiv\mathfrak{S\otimes S},\alpha^{t}\equiv\alpha
\otimes\alpha$ with suitable tensor products $\otimes.$ Further, let $C\left(
\varphi\right)  $ be the complexity of a state $\varphi\in$ $\mathfrak{S}$ and
$T\left(  \varphi;\Lambda^{*}\right)  $ be the transmitted complexity
associated with the state change$\;\varphi\to\Lambda^{*}\varphi.$ These
complexities $C$ and $T$ are the quantities satisfying the following conditions:

\begin{enumerate}
\item[(i)] For any $\varphi\in\mathfrak{S}$,
\[
C(\varphi)\ge0,\ T(\varphi;\Lambda^{*})\ge0.
\]

\item[(ii)] For any orthogonal bijection $j:ex\mathfrak{S}\mathcal{\rightarrow
}ex\mathfrak{S}$ ( the set of all extreme points in $\mathfrak{S} $ ),
\[
C(j(\varphi))=C(\varphi),
\]
\[
T(j(\varphi);\Lambda^{*})=T(\varphi;\Lambda^{*}).
\]

\item[(iii)] For $\Phi\equiv\varphi\otimes\psi\in\mathfrak{S}_{t}$,
\[
C(\Phi)=C(\varphi)+C(\psi).
\]

\item[(iv)] For any state $\varphi$ and a channel $\Lambda^{*},$%
\[
T(\varphi;\Lambda^{*})\le C(\varphi).
\]

\item[(v)] For the identity map ``id'' from $\mathfrak{S}$ to ${\ }%
\mathfrak{S}$.
\[
T(\varphi;id)=C(\varphi).
\]

\end{enumerate}

\begin{definition}
\noindent:Quantum Information Dynamics (QID) is defined by
\\
$%
\left(  \mathcal{A},\mathfrak{S},\alpha(G);\;\Lambda^{*};\;C(\varphi
),T(\varphi;\;\Lambda^{*})\right) \\
\mathit{and\ some\ relations\ R\ among\ them}\mathrm{.}%
$
\end{definition}

There are several examples of the above complexities $C$ and $T$ such as
quantum entropy and quantum mutual entropy \cite{O1,O6}. Information Dynamics
can be applied to the study of chaos in the following sense:

\begin{definition}
\cite{O7,O9,IKO}$\psi$ is more chaotic than $\varphi$ as seen from the
reference system $\mathcal{S}$ if $C(\psi)\ge C(\varphi)$.
When $\varphi$ changes to $\Lambda^{*}\varphi$, the\textit{\ degree of chaos}
associated to this state change(dynamics) $\Lambda^{*}$ is given by
\[
D(\varphi;\Lambda^{*})=\inf\left\{  \int_{\mathfrak{S}}C(\Lambda^{*}%
\omega)d\mu;\mu\in M\left(  \varphi\right)  \right\}  ,
\]
where $\varphi=\int_{\mathfrak{S}}\omega d\mu$ is a maximal extremal
decomposition of $\varphi$ equipped with a certain topology in the state space
$\mathfrak{S}$ and $M\left(  \varphi\right)  $ is the set of such measures. In
some cases such that $\Lambda^{*}$ is linear, this chaos degree $D(\varphi
;\Lambda^{*})$ can be written as $C(\Lambda^{*}\varphi)-T(\varphi
;\;\Lambda^{*}).$
\end{definition}

Since ID has hierarchy (hierarchical structure), it can be applied several
open systems. Here we apply ID to Brain Dynamics.

\section{Entropic chaos degree}

In the context of information dynamics, a chaos degree associated with a
dynamics in classical systems was introduced in \cite{O7}. It has been applied
to several dynamical maps such logistic map, Baker's transformation and
Tinkerbel map with succesful explainations of their chaotic
characters\cite{IOS}. This chaos degree has several merits compared with usual
measures such as Lyapunov exponent.

Here we discuss the quantum version of the classical chaos degree, which is
defined by quantum entropies in Section 2, and we call the quantum chaos
degree the entropic quantum chaos degree. In order to contain both classical
and quantum cases, we define the entropic chaos degree in C*-algebraic
terninology. This setting will not be used in the sequel application, but for
mathematical completeness we first discuss the C*-algebraic setting.

Let $(\mathcal{A},\mathfrak{S})$ be an input C* system and $(\overline
{\mathcal{A}},\overline{{\mathfrak{S}}})$ be an output C* system; namely,
$\mathcal{A}$ is a C* algebra with unit $I$ and $\mathfrak{S}$ is the set of
all states on $\mathcal{A}$. We assume $\overline{\mathcal{A}}=\mathcal{A}$
for simlicity. For a weak* compact convex subset $\mathcal{S}$ (called the
reference space) of $\mathfrak{S}$, take a state $\varphi$ from the set
$\mathcal{S}$ and let%

\[
\varphi=\int_{\mathcal{S}}\omega d\mu_{\varphi}
\]
be an extremal orthogonal decomposition of $\varphi$ in$\mathcal{\ S}$, which
describes the degree of mixture of $\varphi$ in the reference space
$\mathcal{S}$ \cite{O2,OP}. The measure $\mu_{\varphi}$ is not uniquely
determined unless $\mathcal{S}$ is the Schoque simplex, so that the set of all
such measures is denoted by $M_{\varphi}\left(  \mathcal{S}\right)  .$ The
entropic chaos degree with respect to $\varphi\in\mathcal{S}$ and a channel
$\Lambda^{*}$ is defined by%

\[
D^{\mathcal{S}}\left(  \varphi;\Lambda^{*}\right)  \equiv\inf\left\{
\int_{\mathcal{S}}S^{\mathcal{S}}\left(  \Lambda^{*}\varphi\right)
d\mu_{\varphi};\mu_{\varphi}\in M_{\varphi}\left(  \mathcal{S}\right)
\right\}  \text{ }\left(  3.1\right)
\]
where $S^{\mathcal{S}}\left(  \Lambda^{*}\varphi\right)  $ is the mixing
entropy of a state $\varphi$ in the reference space $\mathcal{S}$
\cite{O4,IKO}. When $\mathcal{S=}\mathfrak{S,}$ $D^{\mathcal{S}}\left(
\varphi;\Lambda^{*}\right)  $ is simply written as$D\left(  \varphi
;\Lambda^{*}\right)  .$ This $D^{\mathcal{S}}\left(  \varphi;\Lambda
^{*}\right)  $ contains both the classical chaos degree and the quantum one.

The claasical entropic chaos degree is the case that $\mathcal{A}$ ia abelian
and $\varphi$ is the probability disribution of a orbit generated by a
dynamics (channel) $\Lambda^{*};$ $\varphi=\sum_{k}p_{k}\delta_{k},$ where
$\delta_{k}$ is the delta measure such as $\delta_{k}\left(  j\right)
\equiv\left\{
\begin{array}
[c]{ll}%
1 & \left(  k=j\right) \\
0 & \left(  k\neq j\right)
\end{array}
\right.  .$ Then the claasical entropic chaos degree is%

\[
D_{c}\left(  \varphi;\Lambda^{*}\right)  =\sum_{k}p_{k}S(\Lambda^{*}\delta
_{k})\text{,}
\]
where $S$ equals to von Neumann's entropy, equivalently Shannon's one\cite{OP}.

We explain the entropic chaos degree of a quantum system described by a
density operator. Let $\digamma^{*}$ be a channel sending a state to a state
and $\rho$ be an intial state. After time $n,$ the state is $\digamma^{*n}%
\rho,$ whose Schatten decomposition is denoted by $\sum_{k}\lambda
_{k}^{\left(  n\right)  }E_{k}^{\left(  n\right)  }.$ Then define a channel
$\Lambda_{m}^{*}$ on $\otimes_{1}^{m}\mathcal{H}$ by%

\[
\Lambda_{m}^{*}\sigma=\digamma^{*}\sigma\otimes\cdots\otimes\digamma
^{*m}\sigma,\text{ }\sigma\in\mathfrak{S}(\mathcal{H}),
\]
from which the entropic chaos degree $\left(  3.1\right)  $ for the channels
$\digamma^{*}$ and $\Lambda_{n}^{*}$ are written as%

\begin{align*}
D_{q}\left(  \rho^{\left(  n\right)  };\digamma^{*}\right)   &  =\inf\left\{
\sum_{k}\lambda_{k}^{\left(  n\right)  }S\left(  \digamma^{*}E_{k}^{\left(
n\right)  }\right)  ;\left\{  E_{k}^{\left(  n\right)  }\right\}  \right\}
,\\
D_{q}\left(  \rho^{\left(  n\right)  };\Lambda_{m}^{*}\right)   &
=\inf\left\{  \frac{1}{m}\sum_{k}\lambda_{k}^{\left(  n\right)  }S\left(
\Lambda_{m}^{*}E_{k}^{\left(  n\right)  }\right)  ;\left\{  E_{k}^{\left(
n\right)  }\right\}  \right\}  ,
\end{align*}
where the supremum is taken over all Schatten decompositions of $\digamma
^{*n}\rho.$

We can judge whether the dynamics $\digamma^{*}$ causes a chaos or not by the
value of D as%

\begin{align*}
D  &  >0\text{ and not constant}\Longleftrightarrow\text{chaotic,}\\
D  &  =\text{constant}\Longleftrightarrow\text{weak stable,}\\
D  &  =\text{0}\Longleftrightarrow\text{stable.}%
\end{align*}

\noindent The classical version of this degree was applied to study the
chaotic behaviors of several nonlinear dynamics \cite{IOS,O7}. The quantum
entropic chaos degree is applied to the analysis of quantum spin
system\cite{IKO2} and quantum Baker's type transformation, and we
could measure the chaos of these systems. The information theoretical meaning
of this degree was explained in \cite{O9,O12}.

\section{Quantum Chaos Degree}

In this section, we apply it to study the appearance of chaos in quantum spin
systems\cite{IKO2}.

The quantum chaos degree has the following properties.

\begin{theorem}
For any $\rho\in\mathfrak{S}\left(  \mathcal{H}\right)  $, $\digamma^{*}:$
$\mathfrak{S}\left(  \mathcal{H}\right)  \rightarrow\mathfrak{S}\left(
\mathcal{H}\right)  $ and $\Lambda_{m}^{*}:$ $\mathfrak{S}\left(  \otimes
_{1}^{m}\mathcal{H}\right)  \rightarrow\mathfrak{S}\left(  \otimes_{1}%
^{m}\mathcal{H}\right)  $ defined above, we have:
\end{theorem}

\begin{itemize}
\item[(1)] Let $U_{t}$ be an unitary operator satisfying $U_{t}=\exp\left(
itH\right)  $ for any $t\in\mathbf{R}$.

If $\digamma^{*}\rho=AdU_{t}\left(  \rho\right)  \equiv U_{t}\rho U_{t}^{*}$ ,
$D_{q}\left(  \rho;\digamma^{*}\right)  =D_{q}\left(  \rho;\Lambda_{m}%
^{*}\right)  =0.$

\item[(2)] Let $\rho_{0}$ be a fixed state on $\mathcal{H}$. If $\digamma
^{*}\rho=\rho_{0}$, $D_{q}\left(  \rho;\digamma^{*}\right)  =D_{q}\left(
\rho;\Lambda_{m}^{*}\right)  =S\left(  \rho_{0}\right)  $.

\item[(3)] Let $\lambda$ be a fixed positive real number. If $\digamma^{*}%
\rho=e^{-\lambda}\rho+\left(  1-e^{-\lambda}\right)  \rho_{0}$, $D_{q}\left(
\rho;\digamma^{*}\right)  =D_{q}\left(  \rho;\Lambda_{m}^{*}\right)  =S\left(
\rho_{0}\right)  .$

\item[(4)] Let $\left\{  P_{n}\right\}  $ be the positive operated measure and
$\digamma^{*}\rho=\sum_{k}P_{k}\rho P_{k}$. Then $D_{q}\left(  \rho
;\digamma^{*}\right)  =D_{q}\left(  \rho;\Lambda_{m}^{*}\right)  =$constant,
that is, $\digamma^{*}$ is weak stable. Moreover, if $\left[  P_{k}%
,\rho\right]  =0$, then $D_{q}\left(  \rho;\digamma^{*}\right)  =D_{q}\left(
\rho;\Lambda_{m}^{*}\right)  =0$,that is, $\digamma^{*}$ is stable.
\end{itemize}

The proof of this theorem is given in \cite{IKO2}. We will apply the quantum
entropic chaos degree to spin 1/2 system. See \cite{IKO2} again for the details.

\bigskip Let $\vec{X}=\left(  x_{1},x_{2},x_{3}\right)  $ be a vector in
$R^{3} $ satisfying $\left\|  \vec{X}\right\|  =\sqrt{\sum_{i=1}^{3}x_{i}^{2}%
}$ $\leq1$ and $I$ be the identity $2\times2$ matrix. Any state $\rho$ in a
spin 1/2 system is expressed as%

\[
\rho=\frac{1}{2}\left(  I+\vec{\sigma}\cdot\vec{X}\right)  ,
\]

\noindent where $\vec{\sigma}=\left(  \sigma_{1},\sigma_{2},\sigma_{3}\right)
$ is the Pauli spin matrix vector;%

\[
\sigma_{1}=\left(
\begin{array}
[c]{ll}%
0 & 1\\
1 & 0
\end{array}
\right)  ,\sigma_{2}=\left(
\begin{array}
[c]{ll}%
0 & -i\\
i & 0
\end{array}
\right)  ,\sigma_{3}=\left(
\begin{array}
[c]{ll}%
1 & 0\\
0 & -1
\end{array}
\right)  .
\]

Let $f$ be a non-linear map from $\mathbf{R}^{3}$ to $\mathbf{R}^{3}$
satisfying $\left\|  f\left(  \vec{X}\right)  \right\|  \leq1$ for any
$\vec{X} \in\mathbf{R}^{3}$ with $\left\|  \vec{X}\right\|  \leq1$. A channel
$\digamma^{*}$ is defined by%

\[
\digamma^{*}\rho=\frac{1}{2}\left(  I+\vec{\sigma}\cdot f\left(  \vec{X}
\right)  \right)
\]

\noindent for any state $\rho$.

We now define Baker's type map and see whether this map produces the chaos.
For any vector\textbf{\ }$\vec{X}=\left(  x_{1},x_{2},x_{3}\right)  \ $ on
$\mathbf{R}^{3}$, we consider the following map $f:\mathbf{R}^{3}%
\rightarrow\mathbf{R}^{3}$ ;%

\[
f\left(  x_{1},x_{2},x_{3}\right)  =\left\{
\begin{array}
[c]{l}%
f_{1}\left(  x_{1},x_{2},x_{3}\right)  \text{ }\left(  -\frac{1}{\sqrt{2}}\leq
x_{1}<0\right) \\
f_{2}\left(  x_{1},x_{2},x_{3}\right)  \text{ }\left(  0\leq x_{1}\leq\frac
{1}{\sqrt{2}}\right)
\end{array}
,\right.
\]
where \hspace{13cm}%

\begin{align*}
f_{1}\left(  x_{1},x_{2},x_{3}\right)   &  =\left(  2a\left(  x_{1}+\frac
{1}{\sqrt{ 2}}\right)  -\frac{1}{\sqrt{2}},\frac{1}{2}a\left(  x_{2}+\frac
{1}{\sqrt{2}} \right)  -\frac{1}{\sqrt{2}},0\right) \\
f_{2}\left(  x_{1},x_{2},x_{3}\right)   &  =\left(  2a\left(  x_{1}+\frac
{1}{\sqrt{ 2}}\right)  -\sqrt{2}a-\frac{1}{\sqrt{2}},\frac{1}{2}a\left(
x_{2}+\frac{1}{\sqrt{2}}\right)  +\frac{1}{\sqrt{2}}a-\frac{1}{\sqrt{2}%
},0\right)  .
\end{align*}

\noindent Whenever $\frac{1}{\sqrt{2}}\leq\left|  x_{1}\right|  \leq1$
$\left(  resp.\frac{1}{\sqrt{2}}\leq\left|  x_{2}\right|  \leq1\right)  $, we
replace $x_{1}$with $0\left(  resp.x_{2}=0\right)  $.

The entropic chaos degree $D\left(  \rho^{\left(  n\right)  };\digamma
^{*}\right)  $ and $D\left(  \rho^{\left(  n\right)  };\Lambda_{m}^{*}\right)
$ can be computed \cite{IKO2}, and the result of $D\left(  \rho^{\left(
n\right)  };\Lambda_{m}^{*}\right)  $ is shown in Fig. 4.1 for an initial
value $\vec{X}=\left(  0.3,0.3,0.3\right)  $. We took $740$ different $a$'s
between 0 and 1 with $m=1000,n=2000$.

\begin{center}
Figure 4.1: The change of $D\left(  \rho^{\left(  n\right)  };\Lambda_{m}%
^{*}\right)  $ w.r.t. $a$
\end{center}

The result shows that the quantum dynamica constructed by Baker's type
transformation is stable in $0$ $\leq a\leq0.5$ and chaotic in $0.5$
$<a\leq1.0$. Though there are several approaches to study chaotic behaviors of
quantum systems, we used a new quantity to measure the degree of chaos for a
quantum system. Our chaos degree has the following merits:(1) once the channel
$\Lambda^{*}$, describing the dynamics of a quantum system, is given, it is
easy to compute this degree numerically; (2) the argorithm computing the
degree is easily set for any quantum state.

\section{Quantum Information Dynamic Description of Brain}

The complexity and the chaos degree can be used to examine the chaotic aspects
of not only several nonlinear classical and quantum physical physics but also
life sciences. We will construct a model describimg the function of brain in
the context of QID.

The brain system $\emph{BS}$ =$\mathfrak{X}$ is supposed to be described by a
triple ( $B(\mathcal{H)}$, $\mathfrak{S}(\mathcal{H)}$ , $\Lambda(G)$ ) on a
certain Hilbert space $\mathcal{H}$.

Further we assume the followings:

(1)$\emph{BS}$ is described by a quantum state and the brain itself is divided
into several parts, each of which corresponds to a Hilbert space , so that
$\mathcal{H}$ =$\oplus_{k}\mathcal{H}_{k}$ and $\varphi=\oplus_{k}%
\mathcal{\varphi}_{k},$ $\mathcal{\varphi}_{k}\in\mathfrak{S}(\mathcal{H}%
_{k}\mathcal{)}$.

(2) Each $\mathcal{\varphi}_{k}$ is an entangled state.

(3) The function (action) of is described by a channel $\Lambda^{*}$%
=$\oplus_{k}\Lambda_{k}^{*}$.

(4) $\emph{BS}$ is composed of two parts; information processing part ''$P$''
and others ''$O$'' (consciousness, memory) so that $\mathfrak{X=}\mathfrak{X}
_{P}\otimes\mathfrak{X}_{O}$, $\mathcal{H}$ =$\mathcal{H}_{P}\mathcal{\otimes
H}_{O}$.

Thus in our model, the brain may be considered as a parallel quantum computer.
We will briefly explain mathematical structure of our model.

Let be $s=\left\{  s^{_{1}},s^{_{2}},\cdots,s^{n}\right\}  $ a given (input)
signal (perception) and be $\overline{s}=\left\{  \overline{s}^{_{1}%
},\overline{s}^{_{2}},\cdots,\overline{s}^{n}\right\}  $ the output signal.
After the signal $s$ enters in the brain, each element $s^{j}$ of $s$ is coded
into a proper quantum state $\varphi^{_{j}}\in\mathcal{H}_{P},$ so that the
state corresponding the signal $s$ is $\varphi=$ $\sum_{j}\lambda_{j}%
\varphi^{_{j}}$ with some suitable weight $\left\{  \lambda_{j}\right\}  .$
This state may be regarded as a state processed by the brain and it is coupled
to a state stored $\varphi_{O}$ as a memory (pre-conciousness) in the brain.
The processing in the brain is expressed by a quantum channel $\Lambda^{\ast}$
(or $\Lambda_{P}^{\ast}\otimes$ $\Lambda_{O}^{\ast})$ properly chosen by the
form a state $\varphi$ entering the brain, through which the outcome becomes
$\Lambda^{\ast}\varphi\otimes$ $\varphi_{O}\equiv\overline{\varphi}$ (it may
have different symmetry to $\varphi).$ The channel is determined by the form
of the network of neurons and some other biochemical actions, and its fuction
is like a (quantum) gate in computer. The outcome state $\overline{\varphi}$
contacts with an operator $Q$ describing the work as noema of consciousness,
after the contact a certain reduction of state is occured, which corresponds
to the noesis of consciousness. A part of the reduced state is stored in the
brain as a memory. The scheme of our model is reprented in the following
Figure 5.1.%

\begin{center}
\includegraphics[
height=4.8611in,
width=5.0142in
]%
{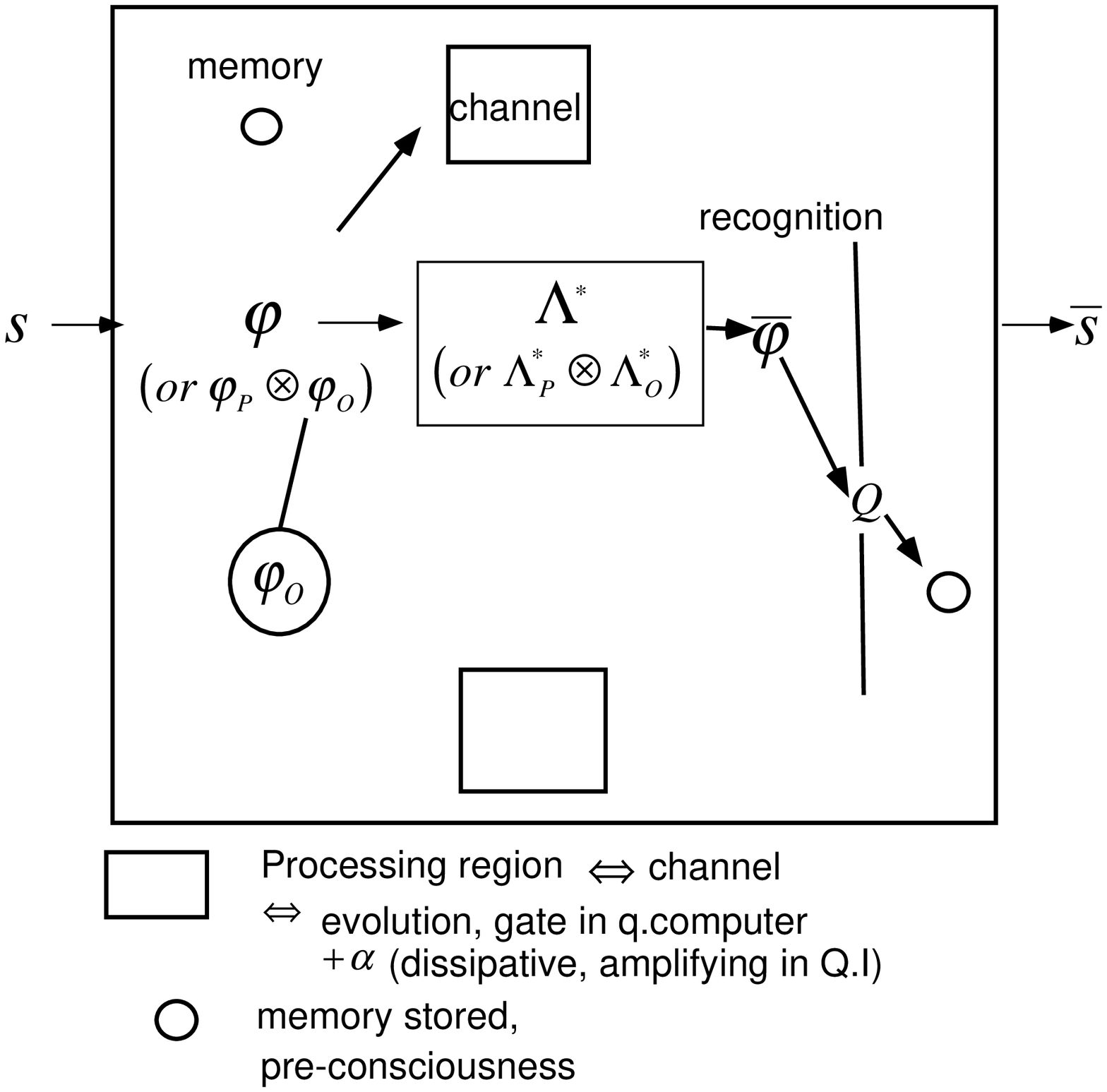}%
\end{center}
%

\[
Fig.5.1
\]

The complex system responses to the information and has a particular role to
choose the information (value of information). That value should be estimated
by a function of the state $\varphi\otimes\varphi_{O},$ the channel
$\Lambda^{*}$ and the operator $Q,$ so that the function $V(\varphi
\otimes\varphi_{O},\Lambda^{*},Q)$ estimating the effect of a signal and a
function of brain is defined as follows:

\begin{definition}
Value of Information and Function:
\end{definition}

(1) $s=\left\{  s^{_{1}},s^{_{2}},\cdots,s^{n}\right\}  $ is more valuable
than $s^{^{\prime}}=\left\{  s^{\prime_{1}},s^{^{\prime}2},\cdots,s^{^{\prime
}n}\right\}  $for $\Lambda^{*}\ $and $Q$ iff
\[
V(^{*}\varphi\otimes\varphi_{O},\Lambda^{*},Q)\geqq V(^{*}\varphi^{^{\prime}%
}\otimes\varphi_{O}^{^{\prime}},\Lambda^{*},Q).
\]

(2) $\Lambda^{*}$ is more valuable than $\Lambda^{^{\prime}*}$ for given
$s=\left\{  s^{_{1}},s^{_{2}},\cdots,s^{n}\right\}  $ and $Q$ iff
\[
V(^{*}\varphi\otimes\varphi_{O},\Lambda^{*},Q)\geqq V(^{*}\varphi
\otimes\varphi_{O},\Lambda^{^{\prime}*},Q).
\]

The details of this estimator is discussed in \cite{O10}, where there exist
some relations between the information of value and the complexity or the
chaos degree.

\end{document}